\documentclass[amsmath,reprint,pre]{revtex4-1}
\usepackage[english]{babel}
\usepackage{graphicx, amsfonts, amssymb, amsmath, nccmath}
\usepackage{hyperref}
\usepackage{colortbl}
\usepackage{siunitx}

\newcommand{\eqr}[1]{Eq.~\eqref{#1}}
\newcommand{\cit}[1]{Ref.~\cite{#1}}
\newcommand{\fig}[1]{Fig.~\ref{#1}}

\newcommand{\nn}{\nonumber\\}

\newcommand{\kbt}{k_{\rm B}T}

\newcommand{\ld}{\lambda_{\mathrm{D}}}
\newcommand{\upd}{\mathrm{d}}

\begin{document}

\author{Mathijs Janssen}
\email{mjanssen@is.mpg.de}
\affiliation{Max-Planck-Institut f\"{u}r Intelligente Systeme, Heisenbergstra{\ss}e 3, 70569 Stuttgart, Germany}
\affiliation{Institut f\"{u}r Theoretische Physik IV, Universit\"{a}t Stuttgart, Pfaffenwaldring 57, 70569 Stuttgart, Germany}
\date{\today}

\begin{abstract}
With two minimal models, I study how electrode curvature affects the response of electrolytes to applied electrostatic potentials. 
For flat electrodes, Bazant \textit{et al.} [\href{http://dx.doi.org/10.1103/PhysRevE.70.021506}{Phys. Rev. E. \textbf{70}, 021506 (2004)}] popularized the $RC$ timescale  $\ld L/D$, with $\ld$ being the Debye length, $2L$ the electrode separation, and $D$ the ionic diffusivity. 
For thin electric double layers near concentric spherical and coaxial cylindrical electrodes, I show here that equivalent circuit models again predict the correct ionic relaxation timescales. 
Importantly, these timescales explicitly depend on both electrode radii, not simply on their difference.
\end{abstract} 

\title{Curvature affects electrolyte relaxation: Studies of spherical and cylindrical electrodes}\maketitle

\section{Introduction}
Many functionalities in nature and technology rely on the out-of-equilibrium behavior of electrolytes.
Transport of ions through nerve membranes, for example,  underlies the firing of neurons \cite{doi:10.1113/jphysiol.1952.sp004764}. 
Similarly, ionic fluxes in nanoporous carbon electrodes determine the power of supercapacitors \cite{beguin2014carbons} and the operation speed of capacitive deionization devices \cite{C5EE00519A}.
To optimize the performance of both capacitive devices through rational design, 
one needs a fundamental understanding of what sets the characteristic timescale $\tau$ of ionic response to electrode potentials.
Since ions forming the electric double layer (EDL) must be partially drawn from a reservoir, 
$\tau$ could depend on ``long" length scales like the electrode separation. 
This makes predicting $\tau$ with molecular simulations difficult, as typical simulation domains only 
capture a small portion of the nanoporous electrode structure \cite{doi:10.1021/nn4058243} or rely on simplified \mbox{geometries \cite{breitsprecher2018charge}}.

Theoretical predictions for  $\tau$ typically concern either nontrivial electrode morphologies, treated approximately \cite{DELEVIE1963751,PhysRevE.81.031502}, or concern
the simplest of geometries, i.e., electrolytes between parallel planar blocking electrodes. 
In the latter case, the ionic charge density reacts to small suddenly imposed (DC) electrostatic potential differences on the timescale $ \ld L/D$, 
which was derived with both microscopic and equivalent circuit model calculations \cite{bazant2004diffuse, janssen2018transient}. 
The bulk diffusion timescale $L^2/D$ can also appear, for instance, when large potentials are applied \cite{bazant2004diffuse} or when the ionic diffusivities are unequal \cite{alexe2007relaxation, balu2018role}. 
Different timescales even (with fractional powers of $\ld$ and $L$) appear when nonblocking electrodes are driven with an AC voltage \cite{PhysRevE.79.021506}.
The analytical parallel-plate results of Refs.~\cite{bazant2004diffuse,  janssen2018transient, alexe2007relaxation, balu2018role, PhysRevE.79.021506} suggest that, to find $\tau$ for capacitive devices with complex nanoporous electrodes, one should identify relevant length scales and their relative importance.
However, there are no general principles yet on how one should go about this task.
Hence, it is timely to diminish the gap between analytical and molecular simulation predictions of electrolyte relaxation.

\begin{figure}[b]
\includegraphics[width=8.6cm]{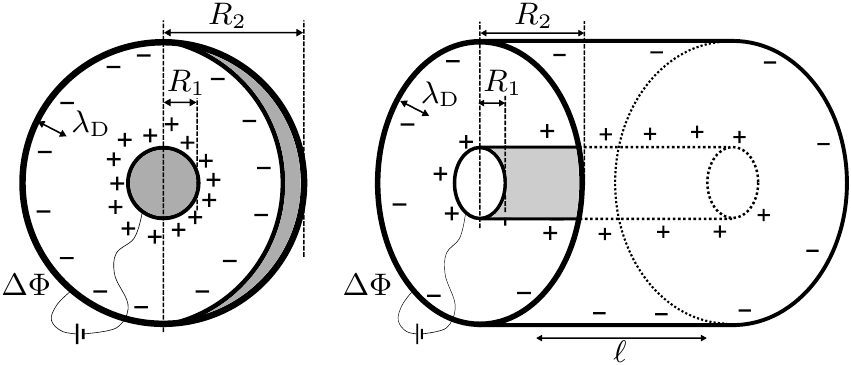}
\caption{Two model EDL capacitors consisting of 1:1 electrolytes with Debye length $\ld$ (solvent and surface charge not shown) between two concentric spherical or coaxial cylindrical electrodes of radii $R_{1}$ and $R_{2}$. At time $t=0$, 
a dimensionless potential difference $\Delta \Phi$ is applied between the electrodes.}
\label{fig1}
\end{figure} 

As a step towards an analytical understanding of the influence of nontrivial electrode morphology on ionic relaxation, in this article 
I discuss EDL capacitors with blocking concentric spherical or coaxial cylindrical  electrodes (see \fig{fig1}).
I use superscript s and c throughout this article to specify observables to either geometry.
For both systems, the electrodes have radii $R_{1}$ and $R_{2}$ ($\Delta R=R_{2}-R_{1}>0$), respectively.
The length $\ell$ of the cylinders is sufficiently large that I can ignore edge effects.
While both systems then contain one relevant geometric length scale more than the parallel plate geometry (depending only on $L$) I will 
show that they allow for similar analytical insight. 
I assume spherical or axial symmetry in either case. Hence, all observables only depend on the radial distance $r$, with $R_{1}\leqslant r\leqslant R_{2}$. 
In between the electrodes is a dilute 1:1 electrolyte of  dielectric constant $\varepsilon$. 
The ionic charge density, the difference between cationic and anionic densities, 
vanishes throughout the cell initially. 
Application of a small dimensionless potential difference $\Delta \Phi\ll1$ 
(with electrostatic potentials measured in units of the thermal voltage $\kbt/e$, with $\kbt$ 
being the thermal energy and $e$ the proton charge) then drives the formation of EDLs at both electrode surfaces. 
Their equilibrium width is set by the Debye length $\ld=\kappa^{-1}$.  

\section{$RC$ reasoning}
Equivalent circuit representations of both setups in \fig{fig1} contain two capacitors representing the EDLs at both electrode surfaces, a resistor for the electrolytic resistance, and a voltage source, all connected in series.
In the spirit of Helmholtz, I treat the EDLs as dielectric capacitors of  width $\ld$ and permittivity $\varepsilon$. Then, using that the capacitance of a dielectric capacitor of 
two conducting concentric spheres at $r_{1}$ and $r_{2}$ is $C^{\rm s}=4\pi \varepsilon /(1/r_{1}-1/r_{2})$, I find the capacitance of the EDL at the inner electrode ($r_{1}=R_{1}, r_{2}=R_{1}+\ld$) as $C^{\rm s}_{R_{1}}\approx 4\pi \varepsilon R_{1}^{2}/\ld$, 
where I assumed  $\ld/R_{1}\ll1$. Likewise, the EDL at the outer electrode ($r_{1}=R_{2}-\ld, r_{2}=R_{2}$) has a capacitance $C^{\rm s}_{R_{2}}\approx 4\pi \varepsilon R_{2}^{2}/\ld$ if $\ld/R_{2}\ll1$. The two in-series EDLs have a total capacitance
\begin{align}\label{eq:capacitance_spherical_RC}
C^{\rm s}=\dfrac{4\pi\varepsilon}{\ld}\dfrac{1}{1/R_{1}^{2}+1/R_{2}^{2}}\,.
\end{align}
The resistance of the electrolyte is $R^{\rm s}=\rho/[4\pi(1/R_{1}-1/R_{2})]$, with $\rho=\ld^2/(\varepsilon D)$ being its resistivity \cite{bazant2004diffuse}. 
Multiplying $R^{\rm s}=\ld^2\Delta R/(4\pi \varepsilon D R_{1}R_{2})$ by $C^{\rm s}$ then yields the $RC$ time
\begin{equation}\label{eq:RCtime_spherical}
\tau^{\rm s}_{RC}=\frac{\ld R_{2}}{D}\frac{1- R_{1}/R_{2}}{R_{1}/R_{2}+R_{2}/R_{1}}\,.
\end{equation}
For the cylindrical electrode system, starting from $C^{\rm c}=2\pi \epsilon \ell /\ln(r_{2}/r_{1})$ and applying the same steps gives
\begin{align}\label{eq:capacitance_cylindrical_RC}
C^{\rm c}=\dfrac{2\pi\varepsilon \ell}{\ld}\dfrac{1}{1/R_{1}+1/R_{2}}\,.
\end{align}
With the resistance $R^{\rm c}=\rho\ln(R_{2}/R_{1})/(2\pi \ell)$, I now find
\begin{equation}\label{eq:RCtime_cylindrical}
\tau^{\rm c}_{RC}=\frac{\ld R_{2}}{ D}\frac{\ln(R_{2}/R_{1})}{1+R_{2}/R_{1}}\,.
\end{equation}
When $R_{1}\to R_{2}$, the electrodes locally resemble parallel plates and the relaxation times reduce to the familiar $\tau^{\rm c}_{RC}\approx \tau^{\rm s}_{RC}\approx \ld \Delta R/(2D)$.
Conversely, for $R_{1}\ll R_{2}$, $\tau^{\rm s}_{RC} \approx R_{1}\ld/D$: The relaxation then only depends on the shortest geometric length scale.  
However, for general cases, $\tau^{\rm s}_{RC}$ and $\tau^{\rm c}_{RC}$ explicitly depend on both $R_{2}$ \textit{and} $R_{1}$. 

\section{Microscopic model}
\subsection{Governing equations}
The dimensionless electrostatic potential $\phi(r,t)$ is related to the dimensionless ionic charge density $q(r,t)$ via the Poisson equation
\begin{align}\label{eq:Poisson}
\frac{2}{r^{d}}\partial_{r}[r^{d}\partial_{r}\phi]&=-\kappa^{2}q\,,
\end{align}
where ${d}=0$ for rectangular, ${d}=1$ for cylindrical, and ${d}=2$ for spherical coordinates.
Moreover, $q(r,t)$ satisfies a continuity equation $\partial_{t}q=-\nabla\cdot\textbf{J}_{q}$. 
At small potentials $\phi(r)\ll1$, $\textbf{J}_{q}=\hat{e}_{r}J_{q}$ with $J_{q}=-D(\partial_{r}q+2\partial_{r}\phi)$ \cite{bazant2004diffuse, janssen2018transient}.
Inserting $\textbf{J}_{q}$ into the continuity equation and using \eqr{eq:Poisson} yields  the 
Debye-Falkenhagen \mbox{equation \cite{debye1928dispersion}}
\begin{align}\label{eq:DebyeFalkenhagen}
\frac{\partial_{t}q}{D}=-\frac{1}{r^{d}}\partial_{r}[r^{d}\partial_{r}q]+\kappa^2 q\,,
\end{align}
subject to 
\begin{subequations}\label{eq:bcs}
\begin{align}
	q(r,{t}=0)&=0 \,, \label{eq:initialcharge} \\
	\phi({R_{2}},t>0)-\phi({R_{1}},t>0)&=\Delta\Phi \,, \label{eq:potentialdrop}\\
	J_{q}(R_{1},{t})=J_{q}(R_{2},{t})&=0\,,\label{noflux}
\end{align}
\end{subequations}
which account for initial charge neutrality, the suddenly imposed potential difference, and the no-flux (blocking) boundary conditions.

\subsection{Solution to Laplace-transformed Debye-Falkenhagen equation}
I determine $q(r,t)$ as follows. With Laplace transformations, 
the partial differential equation for $q(r,t)$ [\eqr{eq:DebyeFalkenhagen}] turns into a solvable ordinary differential equation [\eqr{eq:DebyeFalkenhagenLaplace}] for its Laplace transformed
counterpart $\hat{q}(r,s)=\mathcal{L}\{q(r,t)\}$ [likewise $\hat{\phi}(r,s)=\mathcal{L}\{\phi(r,t)\}$].
Then $q(r,t)$ is determined through 
\begin{align}\label{eq:qtot}
q(r,t)&=\sum_{j}\text{Res}\left(\hat{q}(r,s)\exp(st),s_{j}\right)\,,
\end{align}
with $s_{j}$ being the poles of $\hat{q}(r,s)$, labeled with $j$.
 
Applying $\mathcal{L}\{\,\}$ on both sides of Eqs.~\eqref{eq:Poisson}--\eqref{eq:bcs}, I find 
\begin{subequations}
\begin{align}
\frac{2}{r^{d}}\partial_{r}\left[r^{d}\partial_{r}\hat{\phi}\right]&=- \kappa^2 \hat{q}\,,\label{eq:PoissonLaplace}\\
\frac{1}{r^{d}}\partial_{r}\left[r^{d}\partial_{r}\hat{q}\right]&=k^2\hat{q}\,,\label{eq:DebyeFalkenhagenLaplace}
\end{align}
\end{subequations}
with $k^2=\kappa^2+s/D$, subject to
\begin{subequations}\label{eq:bcLaplace}
\begin{align}
\hat{\phi}({R_{2}},s)-\hat{\phi}({R_{1}},s)&=\frac{\Delta\Phi }{s}\,, \label{eq:potentialdropLaplace}\\
-\partial_{r} \hat{q}-2\partial_{r} \hat{\phi}\Big|_{r=\{R_{1},R_{2}\}}&=0 \,,\label{eq:currentLaplace}
\end{align}
\end{subequations}
where I used \eqr{eq:initialcharge} for \eqr{eq:DebyeFalkenhagenLaplace}.
The solution to  \eqr{eq:DebyeFalkenhagenLaplace} reads $\hat{q}^{\textrm{s}}(r)=a_{2}\exp{[-kr]}/r+b_{2}\exp{[kr]}/r$ for ${d}=2$ and
$\hat{q}^{\textrm{c}}(r)=a_{1}I_{0}(kr)+b_{1}K_{0}(kr)$ for ${d}=1$, with $I_{0}\!$ and $K_{0}\!$ being modified Bessel functions of the first and second kind, respectively. 
The constants $a_{1}$, $a_{2}$, $b_{1}$, and $b_{2}$ could be fixed with the boundary conditions \eqr{eq:bcLaplace}, which, however, inconveniently contain both $\hat{q}$ and $\hat{\phi}$.
Aiming at two constraints on $\hat{q}$  only,
I integrate \eqr{eq:PoissonLaplace} over $\int_{R_{1}}^{r}\upd r $ and use \eqr{eq:currentLaplace} to find
\begin{align}\label{eq:integral_R1}
-2r^{d}\partial_{r}\hat{\phi}=R_{1}^{d}\partial_{r}\hat{q}(R_{1})+\int_{R_{1}}^{r}\upd r \,r^{d} \kappa^{2} \hat{q}\,,
\end{align}
with $\partial_{r}\hat{q}(R_{1})$ shorthand for $\partial_{r}\hat{q}(r)\big|_{r=R_{1}}$.

Repeating the same calculation for $\int_{R_{2}}^{r}\upd r $ gives
\begin{align}\label{eq:integral_R2}
-2r^{d}\partial_{r}\hat{\phi}=R_{2}^{d}\partial_{r}\hat{q}(R_{2})+\int_{R_{2}}^{r}\upd r \,r^{d} \kappa^{2} \hat{q}\,.
\end{align}
The  difference and the sum (integrated over $\int_{R_{1}}^{R_{2}}\upd r$) of Eqs.~\eqref{eq:integral_R1} and \eqref{eq:integral_R2} read
\begin{subequations}
\begin{align}
0&=R_{1}^{d}\partial_{r}\hat{q}(R_{1})-R_{2}^{d}\partial_{r}\hat{q}(R_{2})+\int_{R_{1}}^{R_{2}}\upd r \,r^{d} \kappa^{2} \hat{q}\,,\label{eq:possiblychargeneutrality}\\
-\frac{4\Delta\Phi}{s}&=
\int_{R_{1}}^{R_{2}} \frac{\upd r}{r^{d}}\left[\int_{R_{1}}^{r}\upd r \,r^{d} \kappa^{2} \hat{q}+\int_{R_{2}}^{r}\upd r \,r^{d} \kappa^{2} \hat{q}\right]\hspace{5cm}\nn
&\quad+\left[R_{1}^{d}\partial_{r}\hat{q}(R_{1})+R_{2}^{d}\partial_{r}\hat{q}(R_{2})\right]\int_{R_{1}}^{R_{2}} \frac{\upd r}{r^{d}}\,,\label{eq:constraint2}
\end{align}
\end{subequations}
which are two constraints on $\hat{q}^{\textrm{s}}(r,s)$ and $\hat{q}^{\textrm{c}}(r,s)$ each, which fix the constants $a_{1}$, $a_{2}$, $b_{1}$, and $b_{2}$ therein
\footnote{For ${d}=0$, inserting $\hat{q}(r)=a_{0}\exp{[-kr]}+b_{0}\exp{[kr]}$ gives $\hat{q}(r)$ as reported in  Refs.~\cite{bazant2004diffuse, janssen2018transient}.}.
For spherical electrodes, I find 
\begin{subequations}\label{eq:laplacecharge_spherical}
\begin{flalign}
\hat{q}^{\textrm{s}}&\equiv\frac{2\Delta \Phi}{ s }\frac{ \Gamma^{\textrm{s}}}{\Upsilon^{\textrm{s}}}\,,\label{eq:laplacecharge_spherical_def}\\
\Gamma^{\textrm{s}}&=\big\{m\xi \cosh [m(\xi-\bar{r})]-m \cosh [m(1-\bar{r})]\nn
&\quad-\sinh [m(\xi-\bar{r})]+\sinh [m(1-\bar{r})] \big\}/\bar{r}\,,\label{eq:laplacecharge_spherical_numerator}\\
\Upsilon^{\textrm{s}}&=\frac{2n^{2}}{m}+m\left(  2  -\frac{2n^{2}}{m^{2}} - \xi - \frac{1}{\xi}   \right)\cosh[m(1-\xi)]\nn
&\quad -\left(  m^{2} - n^{2} -\frac{1}{\xi }\right)(1-\xi)\sinh[m(1-\xi)]
 \,,\label{eq:laplacecharge_sphericaldenominator}
\end{flalign}
\end{subequations}
while for cylindrical electrodes I find 
\begin{subequations}\label{eq:laplacecharge_cylindrical}
\begin{align}
&\hat{q}^{\textrm{c}}\equiv \frac{2\Delta \Phi  }{s}\frac{\Gamma^{\textrm{c}}}{\Upsilon^{\textrm{c}}}\label{eq:laplacecharge_cylindrical_def}\,,\\
&\Gamma^{\textrm{c}}= m^2\big\{\left[\xi K_{1}\!\left(m\xi\right) - K_{1}\!\left(m\right)\right]I_{0}\!\left(m\bar{r}\right)\nn
&\quad\quad\quad\quad+ \left[\xi I_{1}\!\left(m\xi\right) -  I_{1}\!\left(m\right)\right] K_{0}\!\left(m\bar{r}\right)\big\}\,,\label{eq:laplacecharge_cylindrical_numerator}\\
&\Upsilon^{\textrm{c}}= m(m^{2}-n^{2}) \xi  \ln{\xi}\left[I_{1}\!\left(m\right)K_{1}\!\left( m\xi\right) -I_{1}\!\left(m\xi\right) K_{1}\!\left( m\right)\right]
\nn
&\quad\quad +\frac{2n^2}{m} -n^2\big[\xi I_{1}\!\left(m\xi\right)K_{0}\!\left(m\right)  +I_{1}\!\left(m\right)K_{0}\!\left(m\xi\right)\nn 
&\quad\quad\quad\quad\quad\quad +\xi I_{0}\!\left(m\right) K_{1}\!\left(m\xi\right)+I_{0}\!\left(m\xi\right) K_{1}\!\left(m\right)\big]
\,,\label{eq:laplacecharge_cylindricaldenominator}
\end{align}
\end{subequations}
where  $m\equiv k R_{2}$,  $n\equiv \kappa R_{2}$, $\xi\equiv R_{1}/R_{2}$, and $\bar{r}=r/R_{2}$.
Here $n$ measures the thickness of the EDLs relative to the system size. For most practical devices, the nanometer-sized EDLs are well separated, i.e., $n\gg1$.

\subsection{Equilibrium}
The pole $s_{0}\equiv0$ in Eqs.~\eqref{eq:laplacecharge_spherical_def} and \eqref{eq:laplacecharge_cylindrical_def} sets 
the equilibrium charge density through $q_{\textrm{eq}}(r)\equiv \text{Res}\left(\hat{q}(r,s),0\right)$. This  amounts to 
$q^{\textrm{s}}_{\textrm{eq}}(r)\equiv2\Delta \Phi \Gamma_{\textrm{eq}}^{\textrm{s}}/\Upsilon_{\textrm{eq}}^{\textrm{s}}$, with
\begin{subequations}\label{APeq:equilcharge_spherical}
\begin{align}
\Gamma_{\textrm{eq}}^{\textrm{s}}&=\big\{n\xi \cosh [n(\xi-\bar{r})]-n \cosh [n(1-\bar{r})]\nn
&\quad-\sinh [n(\xi-\bar{r})]+\sinh [n(1-\bar{r})]\big\}/\bar{r} \,,\\
\Upsilon_{\textrm{eq}}^{\textrm{s}}&=
 \frac{1-\xi}{\xi }\sinh[n(1-\xi)]- n \frac{\xi^2 + 1 }{\xi} \cosh[n(1-\xi)] \nn
 &\quad 
 +2n\,,\label{eq:eqcharge_sphericaldenominator}
\end{align}
\end{subequations}
and $q^{\textrm{c}}_{\textrm{eq}}(r)\equiv2\Delta \Phi \Gamma^{\textrm{c}}_{\textrm{eq}}/\Upsilon^{\textrm{c}}_{\textrm{eq}}$, with
\begin{subequations}\label{APeq:equilcharge_cylindrical}
\begin{align}
\Gamma^{\textrm{c}}_{\textrm{eq}}&= \left[\xi K_{1}\!\left(n\xi\right) - K_{1}\!\left(n\right)\right]I_{0}\!\left(n\bar{r}\right)\nn
&\quad+ \left[\xi I_{1}\!\left(n\xi\right) -  I_{1}\!\left(n\right)\right] K_{0}\!\left(n\bar{r}\right)\,,\\
\Upsilon^{\textrm{c}}_{\textrm{eq}}&= -\xi I_{1}\!\left(n\xi\right)K_{0}\!\left(n\right)  -I_{1}\!\left(n\right)K_{0}\!\left(n\xi\right) \nn
&\quad-\xi I_{0}\!\left(n\right) K_{1}\!\left(n\xi\right)-I_{0}\!\left(n\xi\right) K_{1}\!\left(n\right)+\frac{2}{n}
\,,\label{eq:charge_cylindricaldenominator}
\end{align}
\end{subequations}
respectively [i.e., $\Gamma_{\textrm{eq}}\equiv \Gamma(m=n)$ and $\Upsilon_{\textrm{eq}}\equiv \Upsilon(m=n)$].
Note that $q^{\textrm{s}}_{\textrm{eq}}(r)$ and $q^{\textrm{c}}_{\textrm{eq}}(r)$ can be derived more easily: At equilibrium,
$J_{q}(r)=0$  gives $q(r)=-2\phi(r)$, and thus $\partial_{r}[r^{2}\partial_{r}\phi]=r^{2}\kappa^{2}\phi$ [\eqr{eq:Poisson}]. 
The solution to this equation, $\phi^{\textrm{s}}(r)=A_{2}\exp{[-\kappa r]}/r+B_{2}\exp{[\kappa r]}/r$, contains two constants ($A_{2}$ and $B_{2}$), which are fixed with 
\eqr{eq:potentialdrop} and particle conservation, $\int_{R_{1}}^{R_{2}}\upd r \, r^2 q=0\Rightarrow r^{2}\partial_{r}\phi\big|_{R_{1}}^{R_{2}}=0$. With  $q(r)=-2\phi(r)$,
$q^{\textrm{s}}_{\textrm{eq}}(r)$ then trivially follows. The same steps for $d=1$ lead to $q^{\textrm{c}}_{\textrm{eq}}(r)$.
Further details on the equilibrium EDL near curved electrodes can be found in  Refs.~\cite{lian2016generic, doi:10.1063/1.4979947, doi:10.1021/acs.jpcc.8b10933, asta2019lattice} and in the Appendix, where I confirm Eqs.~\eqref{eq:capacitance_spherical_RC} and \eqref{eq:capacitance_cylindrical_RC} with $q^{\textrm{s}}_{\textrm{eq}}$ and $q^{\textrm{c}}_{\textrm{eq}}$, respectively.

\subsection{Relaxation time} 
For the relaxation of $q(r,t)$, I need to determine 
the locations of the poles $s_{j}\in\mathbb{C}$ of $\hat{q}(r,s)$.
However, instead of immediately focusing on $s_{j}$,
\cit{janssen2018transient} showed for the case of planar electrodes that it is easier to determine the corresponding poles  $m_{j}\in\mathbb{C}$ of $\hat{q}(r,m)$ first.
As in \cit{janssen2018transient}, each $m_{j}$ that I could find was either purely real or purely imaginary (discussed below). Thus, all corresponding $s_{j}=(m_{j}^{2}-n^2)D/R_{2}^{2}$ are real and, as is turns out, $s_{j\neq0}<0$. 
As I am interested in the late-time response of $q(r,t)$, I focus here on $s_{1}$, the pole closest to $s_{0}$, as this pole sets the late-time relaxation time $\tau_{1}=-1/s_{1}$.

\begin{figure}
\includegraphics[width=8.6cm]{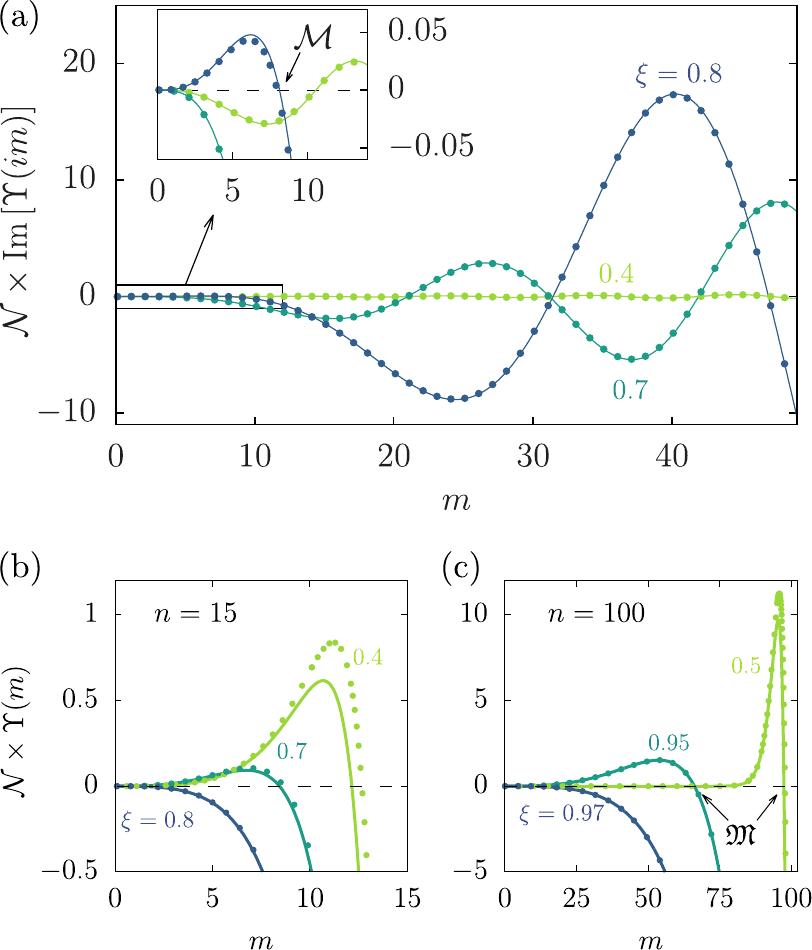}
\caption{Functions $\Upsilon^{\textrm{s}}(m)$ [\eqr{eq:laplacecharge_sphericaldenominator}] (lines) and $\Upsilon^{\textrm{c}}(m)$ [\eqr{eq:laplacecharge_cylindricaldenominator}] (circles), 
both scaled with $\mathcal{N}\equiv\exp{[n(\xi-1)]}$, for several $\xi$ (see labels). (a) Plot of $\textrm{Im}(\Upsilon)$ along the imaginary $m$-axis at $n=15$.
The inset zooms in near $m=0$. 
Also shown is $\Upsilon$ along the real $m$-axis for (b) $n=15$ and (c) $n=100$.}
\label{fig2}
\end{figure}

To find $s_{1}$, I first note that neither $\Gamma^{\textrm{s}}(m)$ nor $\Gamma^{\textrm{c}}(m)$ has poles in $m\in\mathbb{C}$.
Thus, all $s_{j\neq0}$ come
from the zeros of $\Upsilon^{\textrm{s}}(m)$ and $\Upsilon^{\textrm{c}}(m)$, respectively.
Both $\Upsilon^{\textrm{s}}(m)$ and $\Upsilon^{\textrm{c}}(m)$ oscillate around zero on the imaginary $m$-axis [\fig{fig2}(a)] and hence contribute to \eqr{eq:qtot} with infinitely many poles.
However, only the zero at the smallest $m$-value has the potential of leading to $s_{1}$; all zeros further along the imaginary $m$-axis give smaller $s_{j}$, hence faster decaying modes.

Figures~\ref{fig2}(b) and \ref{fig2}(c) show $\Upsilon(m)$ for $m\in\mathbb{R}$ at $n=15$ and $n=100$, respectively.
For certain values of  $\xi$,  
solutions $m_{1}=\mathfrak{M}$ to $\Upsilon^{\textrm{s}}(m)=0$ and  $\Upsilon^{\textrm{c}}(m)=0$  are visible there.
For $n=15$, $\mathfrak{M}$ disappears at $\xi=0.767$ ($\Upsilon^{\textrm{s}}$) and $\xi=0.768$ ($\Upsilon^{\textrm{c}}$), respectively.
Conversely, at $n=100$, $\mathfrak{M}$ persists until higher $\xi$, disappearing only at $\xi=0.965$ (both $\Upsilon^{\textrm{s}}$ and $\Upsilon^{\textrm{c}}$).
In both cases, $m_{1}$ disappears through the origin (not shown), after which it reappears  
as a zero $m_{1}=i \mathcal{M}$ (with $\mathcal{M}\in \mathbb{R}$) on the imaginary $m$-axis [see the inset of \fig{fig2}(b)] that moves away from the 
origin with increasing $\xi$ \footnote{A similar transition from a solution $\mathfrak{M}$ to $\mathcal{M}$ occurs for 
planar electrodes \cite{janssen2018transient}. In that simpler geometry (without $\xi$), the transition happens at $n=\sqrt{3}$ always}. 
(This transition from $\mathfrak{M}$ to $\mathcal{M}$ occurs also at small $\xi$ and $n$ with decreasing $\xi$.)
Associated with these zeros, $s_{1}$ is  either $s_{1}=\left(\mathfrak{M}^2-n^2\right)D/R_{2}^{2}$ or $s_{1}=-\left(\mathcal{M}^2+n^2\right)D/R_{2}^{2}$. 

\begin{figure}[t]
\includegraphics[width=8.6cm]{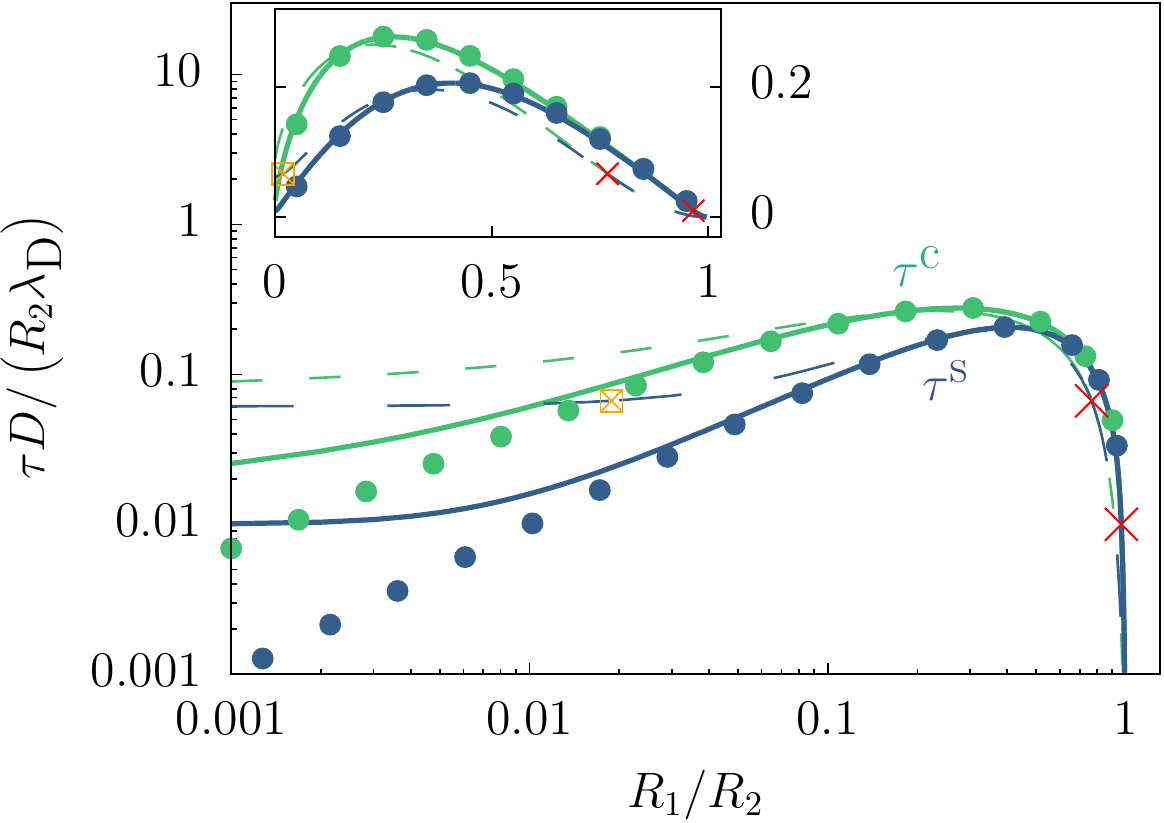}
\caption{Timescales $\tau^{\textrm{s}}_{1}$ (blue) and $\tau^{\textrm{c}}_{1}$ (green) at $\kappa R_{2}=15$ (dashed lines) and $\kappa R_{2}=100$ (solid lines) as 
obtained from numerically solving $\Upsilon^{\textrm{s}}(m)=0$ [\eqr{eq:laplacecharge_sphericaldenominator}] and
$\Upsilon^{\textrm{c}}(m)=0$ [\eqr{eq:laplacecharge_cylindricaldenominator}]. 
Also shown with symbols are $\tau^{\rm s}_{RC}$ [\eqr{eq:RCtime_spherical}] and $\tau^{\rm c}_{RC}$ [\eqr{eq:RCtime_cylindrical}].
The inset plots the same data against linear axes. Red crosses and orange boxed crosses indicate where $m_{1}$ changes from $\mathfrak{M}$ to $\mathcal{M}$ and vice versa, respectively.}
\label{fig3}
\end{figure}  

Figure \ref{fig3} shows $\tau_{1}=-1/s_{1}$ for both setups at $n=\kappa R_{2}=15$ (dashed lines) and $n=100$ (solid lines). 
At red crosses, $m_{1}$ transitions from $\mathfrak{M}$ to $i\mathcal{M}$.
The opposite transition (from $\mathcal{M}$ to $\mathfrak{M}$), indicated with orange boxed crosses, does not occur at $n=100$ and occurs for $\Upsilon^{\textrm{c}}$ only around $\xi=10^{-12}$.
The plateaus at small $\xi$ are understood as follows.
For $\xi\ll1$, $\Upsilon^{\textrm{s}}=0$ reduces to $\tanh m=m$, whose only solution $m_{1}=0$ gives $\tau D/(R_{2}\ld)\approx1/n$. 
Also shown in \fig{fig3} with circles are the equivalent circuit model predictions $\tau^{\rm s}_{RC}$ 
[\eqr{eq:RCtime_spherical}] and $\tau^{\rm c}_{RC}$  [\eqr{eq:RCtime_cylindrical}]. 
For $\xi>0.1$ and $n=15$, $\tau_{1}$ and $\tau_{RC}$ are  qualitatively similar.
For larger $n$, $\tau_{1}$ and $\tau_{RC}$ become identical.
This is understood analytically as follows.
For $m\in \mathbb{R}$ and $n\gg1$, $\Upsilon^{\textrm{s}}(m)=0$ reduces to 
\begin{align}\label{eq:cubiceq}
2\left(m -\frac{n^{2}}{m}\right)&= \left(  m^{2} - n^{2} -\frac{1}{\xi }\right)(1-\xi)+ m \left(\xi + \frac{1}{\xi}\right) \nn
 &\quad+O\left\{\frac{n^2}{m}\exp[m(\xi-1)]\right\}\,.
\end{align}
From \fig{fig2}(c) I see that $\mathfrak{M}\approx n$ if $n\gg1$. 
Inserting the approximation $\mathfrak{M}_{\textrm{ap}}=n -\delta $ into \eqr{eq:cubiceq} and keeping terms up to $O(\delta)$, I find
$\delta=(\xi+1/\xi)/[2(1-\xi)]$. 
This reproduces \eqr{eq:RCtime_spherical}: $\tau^{\textrm{s}}_{1}\approx R_{2}\ld(1-\xi)/[D(\xi+1/\xi)]$.

Similarly, for $n\gg1$, 
$\Upsilon^{\textrm{c}}(m)=0$ 
amounts to
$m \left(m^{2}-n^{2}\right)   \xi  \ln{\xi}=n^{2}\left(1+\xi\right)+O\left\{(\exp[2n(\xi-1)]\right\}$.
[I used Hankel's large argument expansions $I_{\alpha}(z)\sim \exp{(z)}/\sqrt{2\pi z}$ and $K_{\alpha}(z)\sim \sqrt{\pi}\exp{(-z)}/\sqrt{2 z}$ here, which
imply $I_{\alpha}(n)\gg I_{\alpha}(n\xi)$, $K_{\alpha}(\xi n)\gg K_{\alpha}(n)$, $I_{1}(n)/ I_{0}(n)\to1$, and $K_{1}(n)/ K_{0}(n)\to1$]. 
Inserting $\mathfrak{M}_{\textrm{ap}}=n-\epsilon$ and keeping terms up to $O(\epsilon)$ yields $\epsilon=-(1+\xi)/(2 \xi  \ln{\xi})$. This gives $\tau^{\textrm{c}}_{1}
\approx -R_{2}\ld \xi\ln\xi/[D(1+\xi)]$, i.e.,  \eqr{eq:RCtime_cylindrical}.

\subsection{Relaxation of $q(r,t)$}
I use that, close to $s_{1}$,  
$\Upsilon(s)\overset{s\to s_{1}}{=} \partial_{m}\Upsilon(m_{1})\times(s-s_{1})R_{2}^{2}/(2m D)$. 
The slowest relaxation mode $q_{1}(r,t)\equiv\text{Res}\left(\hat{q}(r,s)\exp(st),s_{1}\right)$ now amounts to
\begin{align}\label{eq:qlateprofile}
q_{1}(r,t)&=4\Delta \Phi  \frac{m_{1}}{m_{1}^2-n^2}\frac{\Gamma(m_{1})}{\partial_{m}\Upsilon(m_{1})}\exp\left[-t/\tau_{1}\right]\,.
\end{align}
Explicit expressions for $\partial_{m}\Upsilon(m)$ for the respective geometries follow with  Eqs.~\eqref{eq:laplacecharge_sphericaldenominator} 
and \eqref{eq:laplacecharge_cylindricaldenominator}. I find
\begin{align}
 \partial_{m}\Upsilon^{\textrm{s}}(m)&=m (\xi -1) \left[ \frac{1}{\xi}\left(\xi ^2+1\right)+ \frac{2n^2}{m^{2}}  \right] \sinh [m(1-\xi )]\nn
 &\quad + \left[\frac{2 n^2}{m^2}- (\xi -1)^2 (m^{2}-n^{2}) \right] \cosh [m(1-\xi)]\nn
 &\quad - \frac{2 n^2}{m^{2}}\,,
\end{align}
and
\begin{widetext}
\begin{align}
 \partial_{m}\Upsilon^{\textrm{c}}(m)&=  \left(3 m^2-n^2\right)  \xi \ln (\xi ) [I_1(m) K_1(m \xi )-K_1(m) I_1(m \xi )]-\frac{2n^2}{m^2}\nn
 &\quad+ m (m^2-n^2) \xi  \ln (\xi ) [I_0(m) K_1(m \xi )+K_2(m) I_1(m \xi )-\xi K_1(m) I_0(m \xi )-\xi I_1(m) K_2(m \xi )]\nn
 &\quad+\frac{n^2\xi ^2 }{2}\left\{  [K_0(m \xi )+K_2(m \xi )]I_0(m)- [I_0(m \xi )+I_2(m \xi )]K_0(m) \right\}\nn
 &\quad+\frac{n^2}{2}\left\{[K_0(m)+K_2(m)] I_0(m \xi )-[I_0(m)+I_2(m)] K_0(m \xi )\right\}\,.
 \end{align}
\end{widetext}

 Truncating the sum in \eqr{eq:qtot} after $j=1$, I approximate the relaxation of 
$q(r,t)$   by  
\begin{subequations}\label{eq:qrt}
\begin{align}
q^{\textrm{s}}_{\textrm{ap}}(r,t)&=q^{\textrm{s}}_{\textrm{eq}}(r)+q^{\textrm{s}}_{1}(r,t) \,,\label{eq:qsrt}\\
q^{\textrm{c}}_{\textrm{ap}}(r,t)&=q^{\textrm{c}}_{\textrm{eq}}(r)+q^{\textrm{c}}_{1}(r,t)\,. \label{eq:qcrt}
\end{align}
\end{subequations}
\begin{figure}
\includegraphics[width=8.4cm]{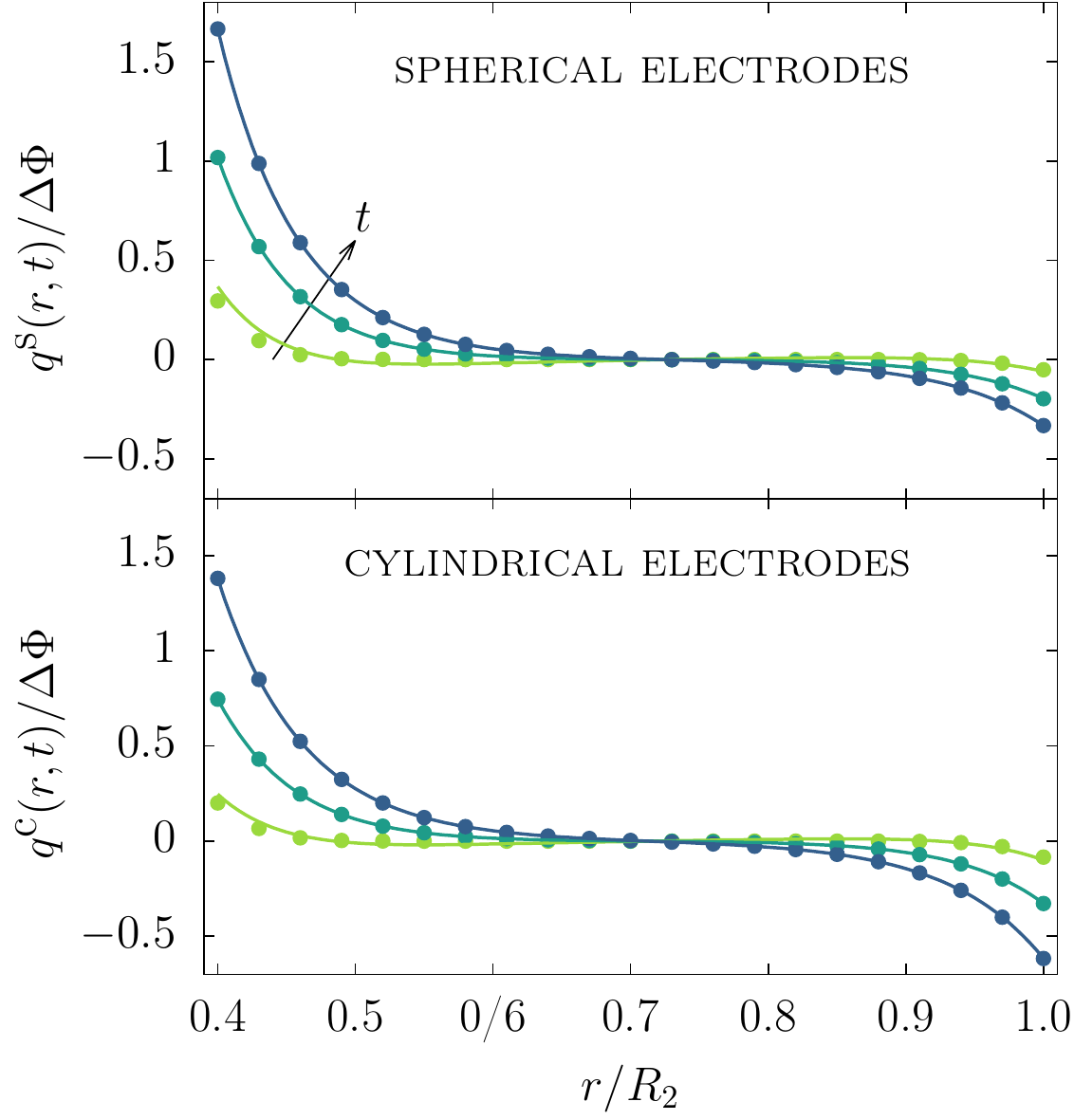}
\caption{Ionic charge densities $q^{\textrm{s}}(r,t)$ (top) and $q^{\textrm{c}}(r,t)$ (bottom), 
for $\kappa R_{2}=15$ and $R_{1}/R_{2}=0.4$ and at times $tD/R_{2}^{2}=\{0.001,0.01,0.1\}$. Shown are 
analytical approximations $q_{\textrm{ap}}(r,t)$ [\eqr{eq:qrt}] (lines) and 
numerical inversions of Eqs.~\eqref{eq:laplacecharge_spherical} and \eqref{eq:laplacecharge_cylindrical}  (circles).}
\label{fig4}
\end{figure} 
In \fig{fig4}, I compare $q_{\textrm{ap}}(r,t)$ 
to numerical inversions 
of Eqs.~\eqref{eq:laplacecharge_spherical} and \eqref{eq:laplacecharge_cylindrical} with the 't Hoog algorithm, respectively. 
I observe a stronger asymmetry in $q^{\textrm{s}}$ than in $q^{\textrm{c}}$, which must stem from the difference in their ratios of inner to outer electrode
surface areas ($\xi^2$ and $\xi$ in either case, respectively).
Note that, at $n=15$ as used here, the numerically determined $\mathfrak{M}^{\textrm{s}}=12.15$  and $\mathfrak{M}^{\textrm{c}}=12.67$
deviate substantially from their analytical approximations $\mathfrak{M}^{\textrm{s}}_{\textrm{ap}}=12.58$ and $\mathfrak{M}^{\textrm{c}}_{\textrm{ap}}=13.10$, respectively;
hence, I use the former.
Clearly, $q^{\textrm{s}}_{\textrm{ap}}(r,t)$ describes $q^{\textrm{s}}(r,t)$ well for
$t\approx \tau^{\textrm{s}}_{1}=0.0129 R^{2}_{2}/D $, while deviations are visible at early times. 
The same is true for $q^{\textrm{c}}_{\textrm{ap}}(r,t)$ and $q^{\textrm{c}}(r,t)$ for times around $\tau^{\textrm{c}}_{1}D/R^{2}_{2}=0.0155$.
Better capturing $q(r,t)$ at early times requires 
truncating the sum in \eqr{eq:qtot} at higher $j$, which I leave for future work. 

Finally, as I focused on electrolytes with equal ionic diffusivities $D_{+}=D_{-}$ in this article, the results derived here can be expected to be accurate for, for instance, KCl, RbBr, and CsBr (which have $D_{+}/D_{-}= 0.97,1.00$, and $1.00$, respectively \cite{agar1989single}), but less so for NaCl ($D_{+}/D_{-}=0.66$). Electrolytes with different ionic diffusivities will probably relax on two timescales: a fast $RC$ timescale as described here and a slower diffusive timescale that becomes more important the more $D_{+}$ and $D_{-}$ differ \cite{alexe2007relaxation, balu2018role, janssen2019transient}.
Finding the precise functional form of this diffusive timescale can be done with calculations along the lines of the ones presented in Refs.~\cite{balu2018role, janssen2019transient}.

\section{Conclusions}
I have studied the influence of electrode morphology on the relaxation of EDL capacitors,
both with equivalent circuit models and with the (microscopic) Debye-Falkenhagen equation.
For two different curved-electrode geometries, I have shown that the timescale of ionic response to an applied electrostatic potential explicitly depends on all geometric length scales present.
The uplifting message is that, for thin EDLs (a case of high practical relevance), easily obtainable $RC$ times capture the ionic relaxation times decently.
Conversely, for thick EDLs, corrections must be taken into account.
These results form a small step towards an analytical understanding of the relaxation of supercapacitors and deionization devices; however, I expect complications at each further step of the way from planar to nanoporous electrodes.

\begin{acknowledgments}
I gratefully acknowledge a discussion with Aymar de Lichtervelde and Pedro de Souza on \cit{de2019heat}, which inspired me to write this article. 
I thank Martin Z. Bazant and Watse Sybesma for insightful comments and I acknowledge S. Dietrich for support.
\end{acknowledgments}

\begin{appendix}

\setcounter{equation}{0}
\renewcommand{\theequation}{{A}\arabic{equation}}
\section{Equilibrium surface charge and capacitance}

The total surface charge $Q^{\textrm{s}}_{1}$ of the smaller electrode is related to its unit surface charge density $\sigma^{\textrm{s}}_{1}$ by 
$Q^{\textrm{s}}_{1}=4 \pi   R_{1}^{2}e \sigma^{\textrm{s}}_{1}$. With Gauss's law 
$e^2 \sigma^{\textrm{s}}_{1}=-\varepsilon \kbt \partial_{r}\phi(R_{1})$ and $\phi(r)=-q(r)/2$  I find $Q_{1}e=2\pi R_{1}^{2}\varepsilon \kbt \partial_{r}q^{\textrm{s}}_{1}(R_{1})$. 
Likewise, the charge on the larger electrode reads $Q^{\textrm{s}}_{2}e=-2\pi R_{2}^{2}\varepsilon \kbt \partial_{r}q^{\textrm{s}}(R_{2})$.
With \eqr{APeq:equilcharge_spherical}, it is now straightforward to verify that the electrodes carry opposite surface charge
$Q^{\textrm{s}}_{1}=-Q^{\textrm{s}}_{2}$.

Likewise, the capacitance $C=eQ_{2}/(\kbt\Delta \Phi)$ reads 
\begin{align}\label{eq:capacitance_spherical}
\frac{C^{\textrm{s}}}{4\pi \epsilon R_{2} }&\overset{\phantom{n\gg1}}{=}-\frac{1}{\Upsilon^{\textrm{s}}_{\textrm{eq}}} \partial_{\bar{r}} \Gamma^{\textrm{s}}_{\textrm{eq}}\big|_{\bar{r}=1}\nn
&\overset{\phantom{n\gg1}}{=} \frac{1}{\Upsilon^{\textrm{s}}_{\textrm{eq}}} \Big\{\left(1-n^2 \xi \right) \sinh [n (1-\xi )]\nn
&\overset{\phantom{n\gg1}}{}\quad\quad\quad+n (\xi -1) \cosh [n (1-\xi)]\Big\}\nn
&\overset{n\gg1}{=}\frac{n\xi}{\xi+1/\xi}+O\left(n^{0}\right)\,.
\end{align}

For the cylindrical electrode system, the surface charge $Q^{\textrm{c}}_{1}=2\pi R_{1}\ell  e \sigma^{\textrm{c}}_{1}$ of the inner electrode
 amounts to $Q^{\textrm{c}}_{1}e=\pi R_{1}\ell  \varepsilon \kbt \partial_{r}q(R_{1})$. 
Similarly, $Q^{\textrm{c}}_{2}e=-\pi R_{2}\ell  \varepsilon \kbt \partial_{r}q(R_{2})$. Again, with \eqr{APeq:equilcharge_cylindrical}, $Q^{\textrm{c}}_{1}=-Q^{\textrm{c}}_{2}$ can be shown to hold.
The capacitance reads
 \begin{align}\label{eq:capacitance_cylindrical}
\frac{C^{\textrm{c}}}{2\pi \epsilon \ell}&\overset{\phantom{n\gg1}}{=} -\frac{1}{\Upsilon^{\textrm{c}}_{\textrm{eq}}} \partial_{\bar{r}} \Gamma^{\textrm{c}}_{\textrm{eq}}\big|_{\bar{r}=\xi}\nn
&\overset{\phantom{n\gg1}}{=}\frac{n\xi}{ \Upsilon^{\textrm{c}}_{\textrm{eq}}}\left[K_{1}\!\left(n\right)I_{1}\!\left(n\xi \right)-I_{1}\!\left(n\right) K_{1}\!(n\xi)\right]\nn
&\overset{n\gg1}{=} \frac{n\xi}{ 1+\xi}+O(\exp{[2n(\xi-1)]})\,,
\end{align}
where, going to the third line, I again used Hankel's large argument expansion, stated below \eqr{eq:cubiceq}.
Note that Eqs.~\eqref{eq:capacitance_spherical} and \eqref{eq:capacitance_cylindrical} are equivalent to Eqs.~\eqref{eq:capacitance_spherical_RC} and \eqref{eq:capacitance_cylindrical_RC}, respectively.
\end{appendix}

\end{document}